\documentclass[fleqn,usenatbib]{mnras}

\usepackage{newtxtext,newtxmath}

\usepackage[T1]{fontenc}
\usepackage{ae,aecompl}

\usepackage{graphicx}	
\usepackage{amsmath}	
\usepackage{amssymb}	
\usepackage{pdflscape}
\usepackage{longtable,booktabs}

\newcommand{\dss} {$\delta$ Scuti stars}
\newcommand{\ds} {$\delta$ Scuti}

\newcommand{\cd} {\mathrm{d}^{-1}}

\newcommand{\mb} {m_{\mathrm{B}}}
\newcommand{\teff} {T_{\mathrm{eff}}}

\newcommand{\numax}{\nu_{\mathrm{max}}}
\newcommand{\amax}{A_{\mathrm{max}}}
\newcommand{\numaxa}{\nu_{\mathrm{max,a}}}
\newcommand{\numaxd}{\nu_{\mathrm{max,d}}}
\newcommand{\numin}{\nu_{\mathrm{min}}}
\newcommand{\nfreq}{N_{\mathrm{f}}}
\newcommand{\effic}{\mathcal{E}}
\newcommand{\Dnu}{\Delta\nu}

\newcommand{\corot} {\emph{CoRoT}}

\newcommand{\kep} {\emph{Kepler}}

\title[Semi-empirical seismic relations of A-F stars]{Semi-empirical seismic relations of A-F stars from CoRoT and Kepler legacy data}
\author[A. Moya et al.]{
A. Moya,$^{1}$\thanks{E-mail: atreyu0@gmail.com (AM)}
J. C. Su\'arez,$^{1}$
A. Garc\'ia Hern\'andez$^{1}$
M. A. Mendoza$^{1}$
\\
$^{1}$University of Granada (UGR). Dept. Theoretical Physics and Cosmology. 18071. Granada. Spain
}

\date{Accepted XXX. Received YYY; in original form ZZZ}

\pubyear{2017}

\begin{document}
\label{firstpage}
\pagerange{\pageref{firstpage}--\pageref{lastpage}}
\maketitle
%
%
\begin{abstract}
Asteroseismology is witnessing a revolution thanks to high-precise asteroseismic space data (MOST, CoRoT, Kepler, BRITE), and their large ground-based follow-up programs. Those instruments have provided an unprecedented large amount of information, which allows us to scrutinize its statistical properties in the quest for hidden relations among pulsational and/or physical observables. This approach might be particularly useful for stars whose pulsation content is difficult to interpret. This is the case of intermediate-mass classical pulsating stars (i.e. $\gamma$ Dor, $\delta$ Scuti, hybrids) for which current theories do not properly predict the observed oscillation spectra. Here we establish a first step in finding such hidden relations from Data Mining techniques for these stars. We searched for those hidden relations in a sample of $\delta$ Scuti and hybrid stars observed by CoRoT and Kepler (74 and 153, respectively). No significant correlations between pairs of observables were found. However, two statistically significant correlations emerged from multivariable correlations in the observed seismic data, which describe the total number of observed frequencies and the largest one, respectively. Moreover, three different sets of stars were found to cluster according to their frequency density distribution. Such sets are in apparent agreement with the asteroseismic properties commonly accepted for A-F pulsating stars.

\end{abstract}




\section{Introduction}\label{sec:intro}

Techniques and tools able to extract semi-empirical relations of observables from the data might be very useful to shed light on any physical phenomena yet poorly understood. In asteroseismology, these relations have been used to derive model-independent fundamental stellar parameters. One of the most illustrative examples comes from the solar-like stars: the progress in the understanding of their internal structure and evolution has been boosted thanks to the scaling relations derived from the large amount of asteroseismic data released by space missions \citep[][and references therein]{Hekker2016}. Such an improvement has permitted the use of these relations as a proxy of masses, radii, mean densities, ages, etc., and thereby to perform more accurate stellar population studies of the galaxy \citep{Miglio2016, Miglio2013}. In the case of $\delta$ Scuti stars, a semi-empirical large separation ($\Delta \nu$) - stellar mean density relation has been derived using pulsating binary stars \citep{GH2015}.

In the era of space asteroseismology, the number of observed stars and the number of detected frequencies has increased by several orders of magnitude \citep{poretti2009,GH2009}. Therefore, data mining statistical tools are now suitable to confirm and fine-tune previously known relations and/or to find new ones. Some of these techniques have been already used in the context of analyzing very large databases of numerical models and observational data \citep{Pichara2016,Verma2016,Guggenberger2016,Hon2017,Angelou2017}.

The case of pulsating A-F stars is especially complex. The presence of a shrinking and very thin outer convective zone and their pulsations around the fundamental radial mode ($\delta$ Scuti) and/or at the asymptotic g-mode regime ($\gamma$ Doradus) make them very unfriendly targets. Since the first space data analyses were done, most of the theoretical support collapsed \citep{Ahmed2010,Uytterhoeven2011,Balona2015}. So far, scholars have tried to understand such a disagreement \citep{Balona2015} without success. Here, we tackle this problem by means of data mining techniques. We seek to find hidden relations among asteroseismic observables, which might help us to better understand the pulsational content of these stars. In the context of $\gamma$ Doradus stars, similar efforts have been undertaken by \citet{vanreeth2015,Alicavus2016,Alicavus2017}. In Section~\ref{sec:data} we describe the data sample used for this study. In section~\ref{sec:patterns} the statistical methods used are explained. Sections~\ref{sec:pairs}  to  \ref{sec:rel2}  are devoted to present the correlations and multi-variable relations found. Section~\ref{sec:fpatterns} discusses the analysis developed for the stellar frequency density functions.  Finally, the conclusions and future prospect are exposed in Section~\ref{sec:conclu}.

\section{Data samples}\label{sec:data}

Since we are interested in extracting information from statistical methods, we need to work with a sufficiently large sample of stars whose light-curves had been observed from space, and for which their physical parameters had been determined. Currently, for A-F pulsating stars, the most complete datasets fulfilling these criteria can be found in the \corot\ \citep{Baglin2006} and \kep\ \citep{Gilliland2010} archives. In the present work, the dataset is composed of  \ds\ and hybrid stars observed by those space missions. This ensures that our results have no instrumental bias or systematic errors.

The \corot\ sample is composed by 90 \dss\ observed on the EXO field during the LRa01 run, i.e., the first long run in the direction of the galactic anti-center, with a time span of 131 days, and a sampling time of 8 minutes. The oscillation spectra were filtered to remove low frequencies below $2\,\cd$. Additional filters were considered in order to avoid spurious peaks. Only those stars for which a periodic pattern is found were kept \citep[see details in][]{Paparo2016}. The final \corot\ sample used in this work has 74 stars with fundamental parameters ($\mathrm{T_{eff}}$, log$g$ and radial velocities), obtained using low-resolution spectroscopic measurements with the AAOmega instrument \citep{Hareter2012}. The absence of frequencies lower than $2\,\cd$ in the \corot\ sample introduces a bias with a low impact on our studies. The analysis of the lower frequency of the sample won't be reliable due to this bias. Nevertheless, the correlations of this observable have been analyzed by \citet{vanreeth2015}. The rest of the observables used in our study are almost insensitive to this filter.

The \kep\ sample contains 153 stars, classified as $\delta$ Scuti or Hybrids following a comprehensive analysis of both their fundamental parameters and their pulsational content \citep[see details in][]{Uytterhoeven2011}. No particular filters were applied to this sample. All these stars were observed in short cadence (58.9s) ensuring that all the frequencies are below the Nyquist frequency.  The only exception is KIC-3429637, observed with the long cadence sampling. This object has the largest amplitude frequency at $10\,\cd$ and a maximum frequency of $22\,\cd$ and therefore only a marginal (if any) contamination by spurious frequencies is expected.  In any case, this particular case remains too marginal to have an impact on the present statistical study.

For the complete sample we extracted the following asteroseismic observables defined as: minimum frequency ($\numin$), maximum frequency ($\numax$), frequency of the mode with the largest amplitude ($\numaxa$), frequency range ($D=\numax$ - $\numin$), number of observed frequencies ($\nfreq$), maximum amplitude ($\amax$), frequency with the largest density of modes ($\numaxd$), and the large separation, $\Dnu$. This latter was obtained using the discrete Fourier Transform to the frequency set \citep[DTFM]{GH2009, GH2013} and the methodology described in \cite{GH2015}. There, the quasi-periodicities found in the observed oscillation spectrum are proved to be large separations in the low order regime\citep[see][for a theoretical support]{Suarez2014}.  However, we could only find the $\Dnu$ in the \kep\ sample for 76 out of the 153 stars.

In addition to those seismic observables, we included the effective temperature ($\teff$), surface gravity ($g$), apparent magnitude at filter "{\it B}" ($\mb$), and two non-compound observables: the convective efficiency of the outer convection zone and the energy of the pulsation mode, defined as $\effic = (\teff^3*\log(g))^{-2/3}$ and $E = [A\cdot \numaxa]^2$ \citep[see details in][]{Uytterhoeven2011}.

We use the apparent magnitude $\mb$ and not the absolute magnitude because we want to incorporate to our study information about our technical capability to detect low amplitude frequencies. For a certain instrument, $\numax$, $\numin$, $D$, and $\nfreq$ can be correlated with the apparent magnitude. On the other hand, the information can offer the absolute magnitude is taken into account using the surface gravity.

In Table \ref{tab:charact} a subset of the complete table is shown. All the data are available on-line.

\section{The method}\label{sec:patterns}

We considered the whole set of observables for the complete set of stars, with no restrictions or conditions of any kind. Since the main goal of the current study is to take a first look at the available data to find possible patterns hidden there, we have used the central observed values without uncertainties. We first analyzed the cross-correlation of all the combinations of pairs of observables, with the aim of getting a first overview of the most direct relations between variables.

With this in mind, for all the observables we performed a systematic search for hidden patterns by applying to each one the following procedure:

\begin{enumerate}
	\item We chose one variable as dependent variable.
	\item We generated all the possible combinations using the remaining '{\it independent}' variables, that is, 4095 combinations.
	\item For every combination, we performed a linear regression with the selected dependent variable (i). This regression provides a list of coefficients (one per variable), their error,  their statistical degree of confidence, and the $R^2$ statistic, interpreted as the variability explained by the regression compared to the total variability of the data.
	\item We selected the regression with the largest $R^2$ value as the reference combination. As expected, we found several relations with a $R^2$ close this reference value. To avoid thus an arbitrary selection, we refined the reference relation. We did so by applying the linear regression explained in the previous step, changing its variables and using the Akaike Information Criterion (AIC) as a tiebreaker \citep{Venables2002}. This provided a final winning combination.
	\item Finally, we studied the relative importance of each variable. Following \cite{Lindeman1980}, we analyzed the covariance matrix to estimate the $R^2$ contribution of each independent variable averaged over orderings among regressors. This allowed us to detect and remove spurious variables from the final selected combination.
\end{enumerate}

Thirteen linear regressions were found, one per observable, although not all of them are statistically significant. Only those combinations explaining at least a 75$\%$ of the variance of the dependent variable, that is, those with $R^2\ge 0.75$, were considered as statistically significant. 

Furthermore, we also searched for non-linear relations. To do so, we applied the procedure described above but with the variables in logarithmic scale. This yields an additional set of thirteen relations, with the same criterion for the statistical significance ($R^2\ge 0.75$).

The difference among $\delta$ Scuti, $\gamma$ Doradus, and hybrid stars is not clearly established. It is usually accepted that frequencies above 5 $\cd$ belong to the $\delta$ Scuti regime and those bellow this limit correspond to the $\gamma$ Doradus regime. However, analyses of space observations yielded frequencies almost everywhere with a significant overlap of both regimes. This contradicts standard non-adiabatic model predictions. In order to find patterns avoiding this uncertainty, we generated an additional set of observables only analyzing the classical $\delta$ Scuti frequency regime. The results found in our study are almost independent of the set considered.

\section{Searching for patterns in the seismic data}

\subsection{Correlation between pairs of observables}\label{sec:pairs}

In order to measure the degree of correlation between pairs of observables, we adopted the Kendall rank correlation coefficient, $\tau$ \citep{Kendall1938}. It measures the ordinal correlation between two variables. That is, we sort and rank the two variables and study the correlation between the positions of every pair of measurements in the respective ranks. Kendall-$\tau$ measures the concordance/discordance probability of the two observables.

\begin{figure}
 \includegraphics[width=\columnwidth]{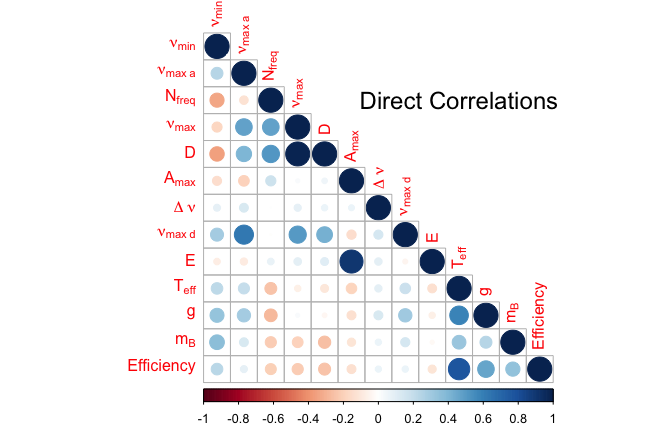}
 \caption{The Kendall-$\tau$ correlation coefficient of all the possible pairs of observables. Color grading from red to blue accounts for anti-correlation to direct correlation, respectively. The larger the circle, the larger the correlation.}
 \label{fig:correl}
\end{figure}

The correlation analysis of all the possible pairs of observables considered in this work shows no clear direct correlations (Fig. \ref{fig:correl}). As expected, the observables constructed as a linear combination of other observables show strong correlations. This is the case of the pairs (D, $\numax$) or (D,  $\numin$) in anti-correlation, to name a few. The rest of the correlations are rather weak. The maximum and minimum frequencies, the frequency of the mode with maximum amplitude, the number of frequencies, and the magnitude, seem to have the larger number of possible correlations although none of them are conclusive. Only $\numaxd$ and $\numaxa$ seem to have a possible although not statistically significant correlation. On the other hand, it is remarkable the lack of correlations of stellar observables such as the effective temperature and the gravity with other variables.

Since no clearly correlated pairs were found, we searched for more complex correlations (next sections).
\begin{table}
 \caption{Coefficients and statistical significance of the first relation, where $\epsilon$ corresponds to the standard error, $t$ represents the $t$-value, $\mathrm{P}$ is P-value, and R.I  represents the relative importance of the variable.}
 \label{tab:coef_rel_1}
 \begin{tabular}{cccccc}
  \hline
  Variable & Coeff. & $\epsilon$ & t  & P & R.I.\\
  \hline
$\log K_1$ & 3.3 & 0.4 & 8.069 & 3.68$\cdot 10^{-13}$ & - \\
$e_1$ ($A_{max}$) & 0.29 & 0.03 & 8.207 & 1.72$\cdot 10^{-13}$ & 0.22\\
$e_2$ ($D$) & 0.84 & 0.09 & 9.021 & 1.80$\cdot 10^{-15}$ & 0.36\\
$e_3$ ($\Delta\nu$) & -0.63 & 0.12 & -5.361 & 3.56$\cdot 10^{-7}$ & 0.05\\
$e_4$ ($m_B$) & -2.4 & 0.3 & -7.898 & 9.41$\cdot 10^{-13}$ & 0.37\\
\hline
 \end{tabular}
\end{table}

\subsection{First relation: Total number of frequencies}\label{sec:rel1}

In order to study non-linear relations, we applied the methodology described in Sect.~\ref{sec:patterns} to the selected variables in logarithmic scale. We found a first relevant relation which links the total number of observed frequencies with the maximum observed amplitude, the stellar magnitude, the range of observed frequencies, and the large separation:
\begin{equation}
\nfreq=K_1 \amax^{e_1}D^{e_2}\Dnu^{e_3}\mb^{e_4},
\label{eq:num_freq}
\end{equation}
where $K_1$ is the constant of proportionality that accounts for the dimensions involved.

The linear regression of this relation using the $\log$ values was performed with 142 stars, i.e. those having all these observables informed. It has a $R^2=0.78$, with a $F-test= 120.2$ with 4 and 133 degrees of freedom, what leads to a P-value smaller than $ 2.2\cdot 10^{-16}$ for the null hypothesis. That is, the regression is statistically significant.

\begin{figure}
 \includegraphics[width=\columnwidth]{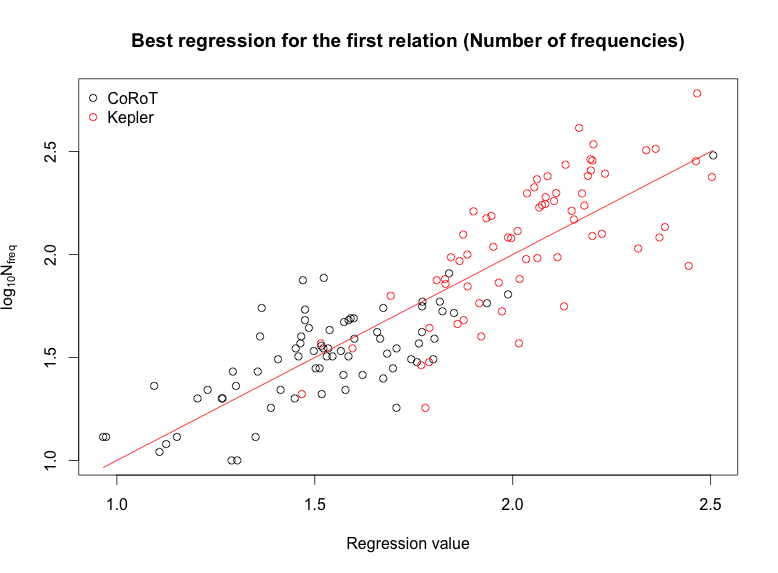}
 \caption{Comparison of the values coming from the regressed relation (abscissa) with the logarithm of the number of frequencies (ordinate). The red line is the 1:1 relation. Black and red circles correspond to stars observed by \corot\ and \kep\ missions, respectively.}
 \label{fig:rel_1_scatter_plot}
\end{figure}

In Table~\ref{tab:coef_rel_1} we show the coefficients obtained for the different independent variables and their uncertainties. In addition, we show the t and P-values of each coefficient to test their statistical significance. Finally, we show the relative importance of each variable as explained in Section~\ref{sec:patterns}.

The statistical significance of every coefficient is really high. Likewise, the magnitude and the frequency range seem to be the most important terms, respectively explaining by themselves a 37$\%$ and 36$\%$ of the 78$\%$ of the variance explained by the regression. On the other hand, the large separation is the term with a smaller weight, at the limit of its rejection as a significant term. Its significant P-value and the decrease in the $R^2$ value of the regression without this term suggest its inclusion.

In Fig.~\ref{fig:rel_1_scatter_plot} we show a comparison of the logarithm of the number of frequencies predicted by the regressed relation and the real logarithm of the number of frequencies. If $R^2=1$ all the points should be placed at the red line. As the regression explains a 78$\%$ of the variance, there is a random distribution around this line.

\begin{figure}
 \includegraphics[width=\columnwidth]{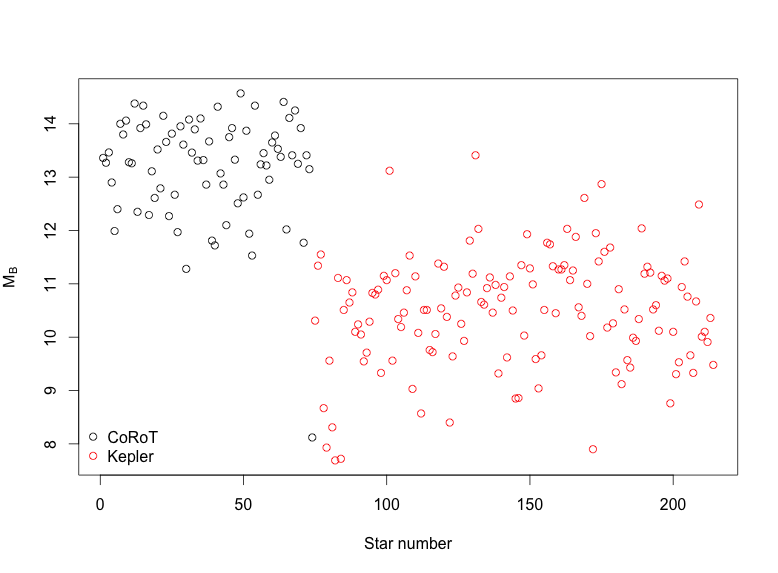}
 \caption{Stellar magnitudes of the sample. Black and red circles correspond to stars observed by \corot\ and \kep\ missions, respectively.}
 \label{fig:magnitudes}
\end{figure}

We see a clear separation between \corot\ and \kep\ samples due to the different magnitudes where these space missions are focused on. While the \corot\ samples are centered around $\mb=13$, the \kep\ ones are centered around $\mb=9.66$ (see Fig.~\ref{fig:magnitudes}).

\subsection{Second relation: Maximum frequency}\label{sec:rel2}

A second relation emerged when examining all the possible correlations among observables. It relates the maximum observed frequency with a set of six observables: $\numaxa$, $\nfreq$, $\Dnu$, $\mb$, $\effic$, and $E$. The observable $\effic$ was re-scaled by 10$^9$ to avoid potential numerical issues. Although those observables present quite different relative importance, the whole set is required to explain the observed variance. The pattern found has the form
\begin{eqnarray}
\numax & = &K_2+c_1\cdot \numaxa+c_2\cdot \nfreq+c_3\cdot \Dnu+ \nonumber \\
&& c_4\cdot \mb+c_5\cdot \effic+c_6\cdot E
\label{eq:freq_max}
\end{eqnarray}
The linear regression of this relation was performed with the subset of 125 stars for which all the selected observables were available. The fit was found statistically significant with $R^2=0.78$, in with a $F-test= 59$ for 6 and 119 degrees of freedom, what leads to a P-value $< 2.2\cdot 10^{-16}$ for the null hypothesis.

\begin{table}
 \caption{Same as Table~\ref{tab:coef_rel_1} but for the second relation found.}
 \label{tab:coef_rel_2}
 \begin{tabular}{cccccc}
  \hline
  Variable & Coeff. & $\epsilon$ & t  & Pr(>|t|) & R.I.\\
  \hline
$K_2$ & 58.54 & 11.56 & 5.064 & 1.52$\cdot 10^{-6}$ & - \\
$c_1$ ($\numaxa$) & 0.88 & 0.01 & 8.929 & 6.37$\cdot 10^{-15}$ & 0.36\\
$c_2$ ($\nfreq$) & 0.081 & 0.014 & 5.850 & 4.41$\cdot 10^{-8}$ & 0.22\\
$c_3$ ($\Delta\nu$) & 1.6 & 0.8 & 2.068 & 0.040801 & 0.13\\
$c_4$ ($m_B$) & -3.2 & 0.7 & -4.389 & 2.48$\cdot 10^{-5}$ & 0.21\\
$c_5$ ($\effic$) & -1.2 & 0.7 & -1.690 & 0.093621 & 0.04\\
$c_6$ ($E$) & 8$\cdot 10^{-5}$ & 2$\cdot 10^{-5}$ & 3.674 & 0.000359 & 0.04\\
\hline
 \end{tabular}
\end{table}

In Table \ref{tab:coef_rel_2} we show the coefficients obtained for the different independent variables and their uncertainties. As in Table \ref{tab:coef_rel_1}, we also show the t and P-values of each coefficient to test their statistical significance and the relative importance of each variable.

The frequency of the maximum amplitude mode justifies itself a 36$\%$ of the 78$\%$ of the variance explained by the regression, thereby becoming the observable with the largest impact on the relation. Moreover, its coefficient is close to the unity, which suggests a significant almost direct relation between $\numax$ and $\numaxa$. The number of frequencies and the magnitude have similar (significant) contributions to the regression (22$\%$ and 21$\%$, respectively). That suggests that, as expected, the instrumental precision plays a key role in finding the maximum frequency. Likewise, the large separation presents a non-negligible relative importance of 13\%.  The remaining $\effic$ and $E$ contribute marginally to the fit. Indeed, for those variables, the t and P-values of the coefficients show that only $c_5$ has poor statistical significance, which, together with their small RI cast doubts on the actual presence of the convective efficiency in the pattern.

\begin{figure}
 \includegraphics[width=\columnwidth]{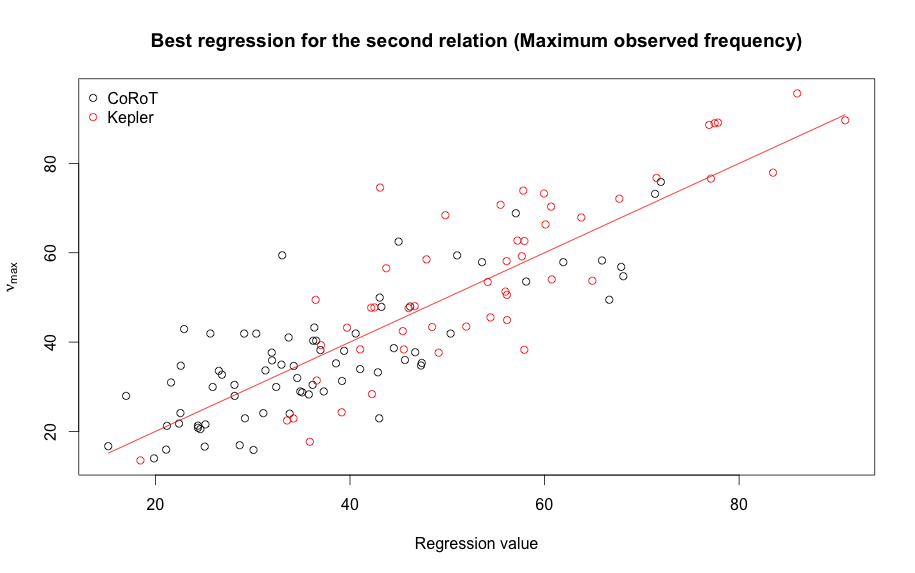}
 \caption{Comparison of the values coming from the regresses relation (abscissa) with the maximum observed frequency (ordinate). The red line is the 1:1 relation. Black and red circles correspond to stars observed by \corot\ and \kep\ missions, respectively.}
 \label{fig:rel_2_scatter_plot}
\end{figure}

In Fig. \ref{fig:rel_2_scatter_plot} we show a comparison of the maximum observed frequency predicted by the regressed relation and the real maximum observed frequency. When $R^2= 1$ all the points should be placed over a line (red line). As the regression explains a 78$\%$ of the variance, there is a random distribution around this line.

\section{Frequency distribution patterns}\label{sec:fpatterns}

Following  \citet{Balona2015},  we sought to extract information from the pulsation frequency distributions. To do so, we analyzed the frequency density of our data samples. For every star, we calculated a Frequency Density Function (FDF) with Gaussian functions as the kernel. The number of Gaussians used to construct the FDF depends on each individual case. No limits for this number were considered.

A huge diversity of  FDF profiles was found with no evident correlations and/or clustering under visual inspection. From this inspection, the commonly accepted transition between $\gamma$ Doradus and $\delta$ Scuti pulsations around $5\,\cd$ is not supported since significant frequency density is found almost everywhere in the frequency spectrum. Therefore, in the case that different clusters with physical information were present, they would overlap at low frequencies and they would have large internal heterogeneity.

We studied potentially hidden clusters from the whole set of calculated FDF in the quest for common properties. To do so, we used the K-means method \citep{lloyd82}. It performs an efficient clustering  \citep{kanungo2002}, especially in the case of complex and overlapped data, based on a simple partition algorithm. After numerous tests, we found 3 as the optimum number of FDF clusters. In this context, optimization is interpreted as the most informative case, i.e. a balance between clusters with different physical properties and noisy results (coming from the overlapping).

\begin{figure}
 \includegraphics[width=\columnwidth]{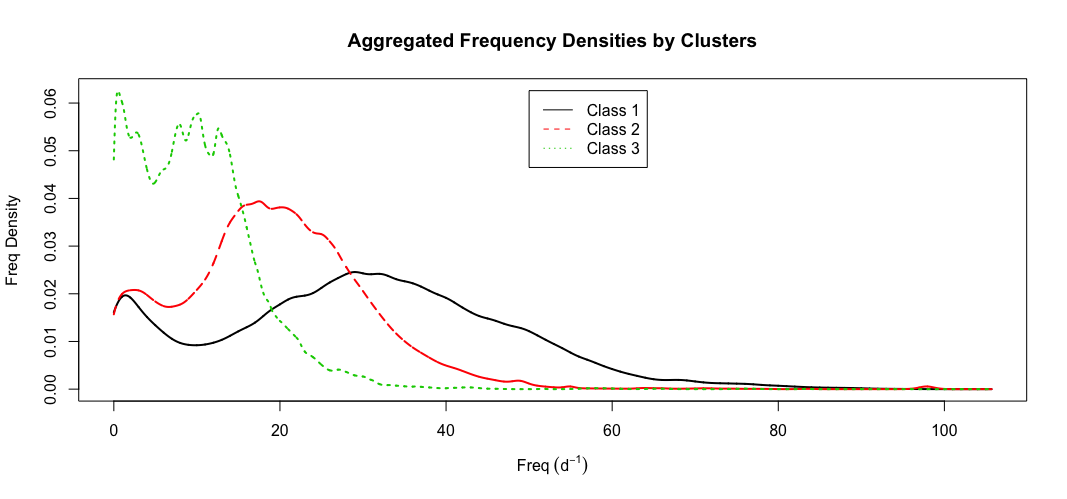}
 \caption{Aggregate FDF of the three clusters found using the K-means algorithm.}
 \label{fig:clusters}
\end{figure}

The analysis of the aggregate FDF for the three clusters (Fig. \ref{fig:clusters} ) revealed some interesting facts:

\begin{enumerate}
\item[1-] All clusters show frequency density at low frequencies.
\item[2-] All clusters exhibit two clearly distinct density regions (at low and high frequencies) that can be regarded as $\gamma$ Doradus and $\delta$ Scuti frequency domains, respectively.
\item[3-] The transition between these two domains presents in all cases a minimum, non-zero, frequency density zone.
\item[4-] In the $\delta$ Scuti domain, the larger the frequency with the maximum density, the larger the dispersion.
\item[5-] The cluster with the $\delta$ Scuti maximum density at lower frequency has the largest $\gamma$ Doradus frequency density.
\end{enumerate}

Inspired by fact (4-), we fitted every FDF to a single Gaussian with the aim of extracting a representative frequency with the maximum density ($\mu$) and a representative dispersion ($\sigma$). We are aware that not all the FDF can be correctly described using a single Gaussian, however, we adopted this hypothesis with the objective of analyzing the general behavior of the clusters.

\begin{figure}
 \includegraphics[width=\columnwidth]{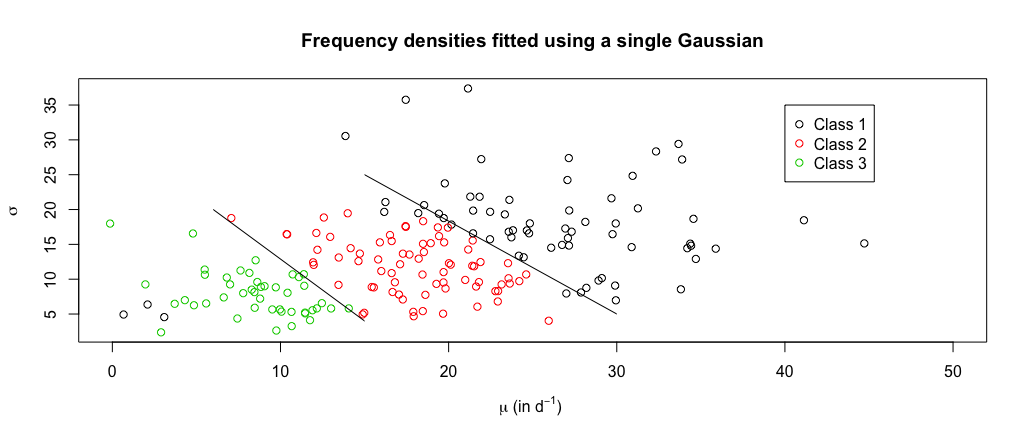}
 \caption{Maximum density ($\mu$) and dispersion ($\sigma$) found fitting each FDF with a single Gaussian. Each cluster is labeled with a different color. The solid lines represent the boundaries between clusters in this parameter space.}
 \label{fig:mu_sigma}
\end{figure}
\begin{figure}
 \includegraphics[width=\columnwidth]{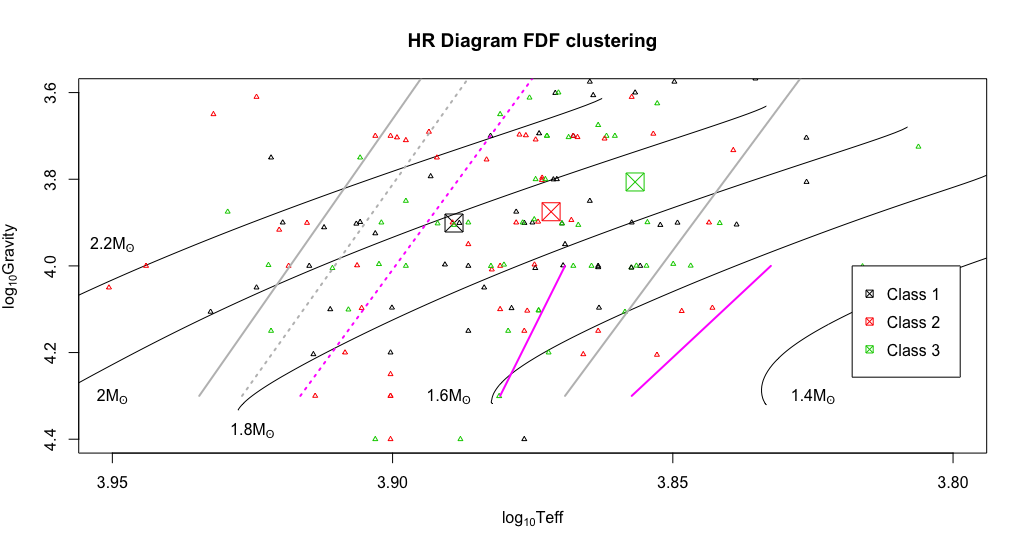}
 \caption{HR-Diagram with the median values of each cluster (crossed-squares). The classical theoretical A-F instability strips are shown in gray ($\delta$ Scuti) and pink ($\gamma$ Doradus) solid lines. Dashed lines are alternative upper limits of the instability strips (see text for details).  The position of every individual star in the sample is included using small empty triangles. The evolutionary tracks of different stellar masses are depicted for reference.}
 \label{fig:HR_FDF}
\end{figure}

When applied this fitting to our data, we obtained a clear separation of the total population into the three sets found by the K-means algorithm in this space of parameters (Fig~\ref{fig:mu_sigma}. This ensures the robustness of the clustering solution found.

\begin{table}
 \caption{Position of the minimum density $\gamma$ Doradus - $\delta$ Scuti transition.}
 \label{tab:transition}
 \begin{tabular}{cccc}
  \hline
  & Class 1 & Class 2  & Class 3\\
  \hline
Frequency (in $d^{-1}$) & 9.83 & 6.62 & 4.74\\
\hline
 \end{tabular}
\end{table}

We studied whether these different groups have different physical characteristics. In our sample, the only non-seismic properties are the effective temperature and the surface gravity, that is, the individual location of each star in the HR-Diagram (Fig. \ref{fig:HR_FDF}). The classical limits of $\delta$ Scuti and $\gamma$ Doradus instability strip were taken, respectively, from \citet{Pamyatnykh2000} and \citet{Dupret2005}. For comparison with recent theoretical developments, we also considered the instability strips predicted by \citet{Xiong2016} (dashed lines at Fig. \ref{fig:HR_FDF}). The low-temperature limits provided by these authors are located bellow the 1.4$M_\odot$ evolutionary track, out of the ranges of our plot. These classical instability zones (especially the $\gamma$ Doradus strip) were no longer valid when space data were analyzed \citep{Uytterhoeven2011,Balona2015}. In Fig. \ref{fig:HR_FDF} we can see how there are stars of every group everywhere, as it has been shown in previous studies.

If we obtain the median of $log_{10}T_{eff}$ and $log_{10}g$ of each cluster, the different groups are distributed in the HR-Diagram as shown in Fig. \ref{fig:HR_FDF}. We can see how these median values recover somehow the classical scheme in $T_{eff}$. In aggregate terms,  class~3 has the largest density at low frequencies. In addition, as the median $T_{eff}$ of the group increases, the position of the $\delta$ Scuti maximum density shifts to larger frequencies, providing a physical explanation for the fact (4-).

Finally, the position of the minimum density $\gamma$ Doradus - $\delta$ Scuti transition also increases with temperature (Table~\ref{tab:transition}). This somehow confirms the widely adopted limit of 5 $\cd$ to discriminate between $\gamma$ Doradus and $\delta$ Scuti frequencies, at least for low-temperature stars. This shows as well, that this transition can be temperature-dependent.

In any case, we must be aware that we are extracting conclusions from median values of a clustering where the members are really heterogeneous. Any application to individual stars must be taken with caution. Therefore these results should be regarded as a first approximation for understanding the underlying physics of the observational data.

\section{Conclusions and future prospects}\label{sec:conclu}

The use of data mining techniques to study a large sample of pulsating A-F stars observed by the CoRoT and Kepler satellites has allowed us to search for relations between seismic magnitudes, which remain hidden from the theoretical analysis. When we analyze correlations between pairs of observables, we don't find any significant result. In addition, when we search for multivariable correlations, two relations were found to be statistically significant, which describe: 

\begin{itemize}
\item The total number of observed frequencies as a non-linear function of the stellar magnitude, the observed frequency range, the maximum amplitude and the large separation, with different relative importances each.

\item The largest observed frequency as a linear function of the frequency of the mode with the largest amplitude, the stellar magnitude, the number of observed frequencies, the large separation and, to a lesser extent, $\effic$ and $E$. Also with different relative importance each.
\end{itemize}

Interestingly, the seismic  components of those relations  (e.g. $\numax$, $\Dnu$, etc. ) are nowadays commonly considered as seismic indices, widely used in solar-like stars\footnote{The Stellar Seismic Indices  (SSI, http://ssi.lesia.obspm.fr/documents/) from the SPACEINN project (http://www.spaceinn.eu) is a good example of the importance of those parameters in Stellar Physics}. This guarantees the availability of such data in the archives for present and future seismic observations \citep[e.g. ][] {Ricker2009, plato}.

Furthermore, a clustering analysis of the sample revealed an interesting classification based on their frequency density distribution. In particular,  low- and medium-frequency domains seem to be separated.  These roughly correspond to the asteroseismic properties accepted for A-F pulsating stars. We also point out to the possible temperature-dependent definition of the transition between $\delta$ Scuti and $\gamma$ Doradus pulsation regimes. This may help to better understand the current open issues regarding the excitation mechanisms of those classical pulsators \citep{Uytterhoeven2011}.

Refinement of the present results will be obtained by increasing the quantity and quality of the sample, which is somehow guaranteed by the new datasets from K2 missions, BRITE, and the coming space missions TESS and PLATO. For the quality, understood as the accuracy in the determination of the different seismic indexes, the future seems promising since a whole community is devoting significant efforts in the scientific preparation of those missions.  

In this context, the present results pave the way to the massive seismic analysis of pulsating A-F stars, and thus better understand their role in the galaxy (e.g. in stellar population studies). To do so, we first need to understand the physics beyond the present results (work in prep.), which will allow us to find proxies for stellar bulk parameters from asteroseismology.

We plan to extend this study to the largest (and heterogeneous) possible sample, using all the new space data catalogs of pulsating A-F stars available  \citep[e.g. ][]{Tkachenko2013,Bradley2015}. The bottle-neck, in this case, is the calculation of the large separation of every star. In that work, such a large sample will enhance the statistics, allowing us to properly assess the impact of the observational uncertainties in our studies. In addition, the publication of new and most accurate physical stellar properties \citep{Huber2014,Alicavus2017} will allow a more reliable conclusion about the impact of $T_{eff}$ and the surface gravity in the relations obtained.

\section*{Acknowledgements}

The authors acknowledge funding support from Spanish public funds for research under project ESP2015-65712-C5-5-R (MINECO/FEDER), and from project RYC-2012-09913 under the 'Ram\'on y Cajal' program of the Spanish MINECO. The authors want to acknowledge the effort and very constructive comments and suggestions of the anonymous referee.

\section*{Software}

Analysis was performed with R version 3.3.1 \citep{R}, RStudio Version 1.0.143, and the R libraries dplyr 0.5.0 \citep{dplyr2016}, MASS 7.3-47 \citep{MASS2017},  relaimpo 2.2-2 \citep{relaimpo2015}, xlsx 0.5.7 \citep{xlsx2014}, foreach 1.4.3 \citep{foreach2015}, fpc \citep{fpc2015}, corrplot 0.86 \citep{corrplot}, and rriskDistributions 2.1.2 \citep{rriskDistributions2017}.




\bibliographystyle{mnras}
\bibliography{patterns_biblio} 
\onecolumn
\begin{landscape}  
 \begin{longtable}{cccccccccccccc}
  \caption{Seismic and physical observables of the stars used.}\label{tab:charact}\\
\toprule
    Star (ID) & $\nu_{min}$ & $\nu_{max}$ & $D=\nu_{max}$ - $\nu_{min}$ & $\nu_{max,a}$ & $N_{freq}$ & $A_{max}$ & $\Delta\nu$ & $\nu_{max,d}$ & $T_{eff}$ & $g$ & $m_B$ & $E$ & $\effic\,(\cdot 10^{-9})$\\
        & $d^{-1}$ & $d^{-1}$ & $d^{-1}$ & $d^{-1}$ & adim. & mmag & $d^{-1}$ & $d^{-1}$ & $^\circ K$ & $m\cdot s^{-2}$ & mag & (mmag$\cdot d^{-1})^2$ & $^\circ K^{-2}$\\
        \midrule\endfirsthead
    \caption*{\tablename{} \ref{tab:charact} (continued)}\\
\toprule
    Star (ID) & $\nu_{min}$ & $\nu_{max}$ & $D=\nu_{max}$ - $\nu_{min}$ & $\nu_{max,a}$ & $N_{freq}$ & $A_{max}$ & $\Delta\nu$ & $\nu_{max,d}$ & $T_{eff}$ & $g$ & $M_B$ & $E$ & $\effic\,(\cdot 10^{-9})$\\ 
            & $d^{-1}$ & $d^{-1}$ & $d^{-1}$ & $d^{-1}$ & adim. & mmag & $d^{-1}$ & $d^{-1}$ & $^\circ K$ & $m\cdot s^{-2}$ & mag & (mmag$\cdot d^{-1})^2$ & $^\circ K^{-2}$\\
            \midrule\endhead
CoRoT-102661211 & 2.95 & 21.75 & 18.81 & 10.02 &  59 & 8.46 & 0.89 & 10.09 & 7075 & 3758.37 & 13.36 & 71.61 & 0.85 \\ 
  CoRoT-102671284 & 1.41 & 29.95 & 28.54 & 9.16 &  42 & 3.55 & 2.14 & 4.19 & 8550 & 4466.84 & 13.27 & 12.59 & 0.58 \\ 
  CoRoT-102702314 & 1.66 & 16.70 & 15.04 & 7.40 &  25 & 7.68 & 0.93 & 8.14 & 7000 & 944.06 & 13.46 & 59.04 & 0.99 \\ 
  CoRoT-102712421 & 17.58 & 43.26 & 25.68 & 21.68 &  32 & 0.85 & 2.29 & 26.15 & 7400 & 8912.51 & 12.90 & 0.73 & 0.73 \\ 
  CoRoT-102723128 & 13.97 & 30.42 & 16.45 & 22.05 &  18 & 0.53 & 1.85 & 25.27 & 6975 & 7943.28 & 11.99 & 0.28 & 0.83 \\ 
  CoRoT-102703251 & 2.21 & 16.60 & 14.38 & 6.41 &  27 & 0.79 & 1.87 & 8.26 & 9100 & 6309.57 & 12.40 & 0.62 & 0.50 \\ 
  CoRoT-102704304 & 3.86 & 27.95 & 24.09 & 7.87 &  53 & 9.72 & 0.78 & 10.30 & 7050 & 1778.28 & 14.00 & 94.48 & 0.92 \\ 
  CoRoT-102694610 & 1.26 & 27.95 & 26.69 & 7.35 &  75 & 3.24 & 4.24 & 9.28 & 8000 & 5011.87 & 13.80 & 10.48 & 0.65 \\ 
  CoRoT-102706800 & 9.51 & 30.96 & 21.45 & 10.92 &  49 & 6.30 & 1.79 & 21.53 & 7125 & 2113.49 & 14.06 & 39.70 & 0.88 \\ 
  CoRoT-102637079 & 10.68 & 33.56 & 22.88 & 13.76 &  43 & 1.49 & 1.37 & 24.97 & 7325 & 7079.46 & 13.28 & 2.23 & 0.76 \\ 
  CoRoT-102687709 & 3.25 & 73.18 & 69.93 & 53.68 &  39 & 2.35 & 3.47 & 53.25 & 7950 & 25118.86 & 13.26 & 5.55 & 0.59 \\ 
  CoRoT-102710813 & 8.81 & 20.52 & 11.71 & 11.40 &  13 & 0.57 & 3.12 & 14.46 & 8350 & 14125.38 & 14.38 & 0.32 & 0.56 \\ 
  CoRoT-102678628 & 0.60 & 23.98 & 23.38 & 14.97 &  77 & 1.37 & 2.81 & 15.56 & 7100 & 1678.80 & 12.35 & 1.87 & 0.91 \\ 
  CoRoT-102599598 & 2.21 & 28.28 & 26.07 & 19.37 &  55 & 1.40 & 3.47 & 19.58 & 7600 & 10000.00 & 13.92 & 1.95 & 0.69 \\ 
  CoRoT-102600012 & 4.70 & 41.91 & 37.20 & 38.34 &  27 & 0.39 & 2.81 & 21.13 & 8000 & 25118.86 & 14.34 & 0.16 & 0.58 \\ 
  CoRoT-102618519 & 14.97 & 59.43 & 44.45 & 18.85 &  54 & 0.61 & 2.23 & 37.15 & 7500 & 31622.78 & 13.99 & 0.38 & 0.65 \\ 
  CoRoT-102580193 & 17.56 & 59.41 & 41.85 & 33.71 &  22 & 0.62 & 3.21 & 34.59 & 7525 & 14125.38 & 12.29 & 0.38 & 0.68 \\ 
  CoRoT-102620865 & 3.02 & 22.94 & 19.91 & 11.01 &  39 & 3.57 & 1.10 & 12.90 & 11250 & 9440.61 & 13.11 & 12.77 & 0.31 \\ 
  CoRoT-102721716 & 3.89 & 57.89 & 54.00 & 35.48 &  52 & 2.48 & 2.43 & 35.33 & 7700 & 14125.38 & 12.61 & 6.13 & 0.65 \\ 
  CoRoT-102622725 & 13.96 & 35.90 & 21.95 & 19.61 &  23 & 0.28 & 4.46 & 24.86 & 6000 & 19952.62 & 13.52 & 0.08 & 1.05 \\ 
  CoRoT-102623864 & 12.07 & 47.96 & 35.89 & 29.83 &  30 & 3.35 & 2.29 & 29.20 & 7900 & 10000.00 & 12.79 & 11.20 & 0.64 \\ 
  CoRoT-102624107 & 3.45 & 34.76 & 31.31 & 34.24 &  32 & 3.13 & 2.10 & 18.81 & 8400 & 11220.18 & 14.15 & 9.82 & 0.56 \\ 
  CoRoT-102724195 & 20.17 & 40.30 & 20.14 & 25.90 &  28 & 1.89 & 1.21 & 26.71 & 7550 & 7943.28 & 13.66 & 3.57 & 0.71 \\ 
  CoRoT-102728240 & 9.73 & 49.96 & 40.24 & 25.84 &  55 & 0.63 & 1.62 & 37.29 & 7450 & 15848.93 & 12.27 & 0.40 & 0.69 \\ 
  CoRoT-102702932 & 2.65 & 21.30 & 18.65 & 15.84 &  48 & 2.75 & 0.81 & 13.51 & 6975 & 2238.72 & 13.81 & 7.58 & 0.92 \\ 
  CoRoT-102603176 & 2.91 & 22.93 & 20.02 & 13.00 &  64 & 23.92 & 0.98 & 11.00 & 12800 & 19952.62 & 12.67 & 572.17 & 0.23 \\ 
  CoRoT-102733521 & 3.22 & 33.66 & 30.44 & 15.28 &  42 & 1.95 & 1.67 & 16.85 & 7125 & 4216.97 & 11.97 & 3.79 & 0.83 \\ 
  CoRoT-102634888 & 3.44 & 34.71 & 31.26 & 12.22 &  39 & 1.87 & 1.34 & 13.84 & 7175 & 10000.00 & 13.95 & 3.51 & 0.77 \\ 
  CoRoT-102636829 & 3.17 & 42.92 & 39.75 & 10.32 &  81 & 4.72 & 1.28 & 10.85 & 7000 & 1584.89 & 13.61 & 22.26 & 0.94 \\ 
  CoRoT-102639464 & 3.03 & 62.49 & 59.46 & 20.23 &  31 & 0.43 & 3.33 & 14.74 & 9450 & 7943.28 & 11.28 & 0.18 & 0.45 \\ 
  CoRoT-102639650 & 13.97 & 47.89 & 33.92 & 30.17 &  28 & 2.29 & 3.39 & 24.14 & 7500 & 7943.28 & 14.08 & 5.27 & 0.72 \\ 
  CoRoT-102641760 & 3.22 & 35.34 & 32.12 & 32.42 &  32 & 1.72 & 2.63 & 17.59 & 7950 & 19952.62 & 13.46 & 2.97 & 0.60 \\ 
  CoRoT-102642516 & 11.94 & 28.95 & 17.01 & 22.20 &  20 & 5.59 & 3.01 & 22.19 & 7275 & 5011.87 & 13.90 & 31.24 & 0.79 \\ 
  CoRoT-102742700 & 9.55 & 33.96 & 24.41 & 24.52 &  28 & 7.88 & 2.40 & 24.78 & 7550 & 7498.94 & 13.31 & 62.13 & 0.71 \\ 
  CoRoT-102743992 & 4.22 & 34.95 & 30.73 & 16.84 &  20 & 0.69 & 4.39 & 30.40 & 7950 & 19952.62 & 14.10 & 0.47 & 0.60 \\ 
  CoRoT-102745499 & 5.55 & 13.97 & 8.42 & 7.27 &  22 & 1.43 & 1.32 & 9.97 & 7900 & 7079.46 & 13.32 & 2.04 & 0.65 \\ 
  CoRoT-102649349 & 2.49 & 27.95 & 25.45 & 12.95 &  16 & 1.14 & 2.12 & 3.21 & 9425 & 8912.51 & -- & 1.31 & 0.45 \\ 
  CoRoT-102647323 & 15.04 & 56.83 & 41.79 & 49.50 &  32 & 0.83 & 3.85 & 45.82 & 8200 & 19952.62 & 12.86 & 0.70 & 0.56 \\ 
  CoRoT-102650434 & 5.86 & 15.93 & 10.07 & 7.30 &  34 & 11.84 & 1.09 & 12.56 & 8500 & 7498.94 & 13.67 & 140.14 & 0.56 \\ 
  CoRoT-102651129 & 13.98 & 49.48 & 35.50 & 45.93 &  35 & 0.44 & 3.52 & 40.24 & 8350 & 5623.41 & 11.81 & 0.20 & 0.59 \\ 
  CoRoT-102753236 & 3.22 & 28.75 & 25.53 & 14.78 &  37 & 3.60 & 2.60 & 14.83 & 7600 & 12589.25 & 11.72 & 12.97 & 0.68 \\ 
  CoRoT-102655408 & 2.80 & 41.90 & 39.11 & 17.80 &  44 & 1.39 & 2.75 & 27.64 & 7375 & 10000.00 & 14.32 & 1.92 & 0.73 \\ 
  CoRoT-102656251 & 6.87 & 31.30 & 24.43 & 25.76 &  22 & 0.54 & 1.62 & 29.04 & 7950 & 15848.93 & 13.07 & 0.29 & 0.61 \\ 
  CoRoT-102657423 & 3.36 & 16.92 & 13.55 & 11.67 &  36 & 5.51 & 2.40 & 9.50 & 8150 & 2660.73 & 12.86 & 30.38 & 0.66 \\ 
  CoRoT-102575808 & 3.15 & 31.96 & 28.80 & 12.92 &  47 & 1.81 & 4.72 & 13.58 & 7250 & 2113.49 & 12.10 & 3.26 & 0.85 \\ 
  CoRoT-102761878 & 0.53 & 15.86 & 15.32 & 15.86 &  11 & 0.77 & 4.31 & 1.89 & 7375 & 5011.87 & 13.75 & 0.59 & 0.77 \\ 
  CoRoT-102576929 & 2.65 & 21.57 & 18.93 & 11.92 &  20 & 0.41 & 1.75 & 13.73 & 8925 & 11220.18 & 13.92 & 0.17 & 0.49 \\ 
  CoRoT-102669422 & 13.97 & 38.04 & 24.07 & 27.88 &  35 & 0.94 & 1.71 & 30.60 & 7300 & 4731.51 & 13.33 & 0.88 & 0.79 \\ 
  CoRoT-102670461 & 7.43 & 33.24 & 25.81 & 28.90 &  49 & 0.87 & 1.28 & 14.99 & 7325 & 3758.37 & 12.51 & 0.76 & 0.80 \\ 
  CoRoT-102607188 & 10.97 & 58.27 & 47.30 & 54.64 &  23 & 0.35 & 3.42 & 49.57 & 8100 & 15848.93 & 14.57 & 0.13 & 0.59 \\ 
  CoRoT-102673795 & 12.97 & 29.95 & 16.98 & 17.37 &  13 & 0.72 & 2.12 & 25.43 & 8050 & 5623.41 & 12.62 & 0.51 & 0.64 \\ 
  CoRoT-102773976 & 12.97 & 35.99 & 23.02 & 32.24 &  13 & 0.10 & 3.73 & 14.67 & 7525 & 25118.86 & 13.87 & 0.01 & 0.66 \\ 
  CoRoT-102775243 & 10.98 & 57.89 & 46.91 & 40.83 &  31 & 2.12 & 3.68 & 42.55 & 7950 & 17782.79 & 11.94 & 4.52 & 0.60 \\ 
  CoRoT-102775698 & 2.83 & 28.95 & 26.12 & 15.24 &  56 & 1.24 & 1.13 & 15.81 & 9550 & 5623.41 & 11.53 & 1.54 & 0.45 \\ 
  CoRoT-102675756 & 5.08 & 20.82 & 15.74 & 14.59 &  40 & 3.28 & 2.14 & 13.99 & 7350 & 1496.24 & 14.34 & 10.73 & 0.86 \\ 
  CoRoT-102677987 & 13.97 & 38.20 & 24.24 & 23.62 &  26 & 0.90 & 1.18 & 27.81 & 7700 & 8912.51 & 12.67 & 0.81 & 0.67 \\ 
  CoRoT-102584233 & 12.97 & 37.64 & 24.68 & 21.15 &  12 & 0.16 & 3.47 & 14.99 & 6400 & 5308.84 & 13.24 & 0.03 & 1.02 \\ 
  CoRoT-102785246 & 13.97 & 34.66 & 20.69 & 21.90 &  37 & 1.62 & 1.76 & 22.03 & 7425 & 6309.57 & 13.45 & 2.62 & 0.74 \\ 
  CoRoT-102686153 & 1.39 & 21.26 & 19.86 & 8.83 &  31 & 1.16 & 2.03 & 3.00 & 7125 & 3349.65 & 13.22 & 1.34 & 0.85 \\ 
  CoRoT-102786753 & 2.52 & 24.11 & 21.59 & 8.92 &  59 & 8.03 & 1.10 & 14.17 & 7100 & 2660.73 & 12.95 & 64.40 & 0.87 \\ 
  CoRoT-102787451 & 1.99 & 24.09 & 22.10 & 17.42 &  13 & 0.37 & 3.68 & 17.13 & 7300 & 10000.00 & 13.65 & 0.14 & 0.74 \\ 
  CoRoT-102587554 & 2.59 & 41.91 & 39.32 & 18.26 &  34 & 0.61 & 1.71 & 18.94 & 7375 & 5011.87 & 13.78 & 0.37 & 0.77 \\ 
  CoRoT-102688156 & 2.51 & 35.25 & 32.74 & 21.79 &  21 & 2.27 & 4.03 & 3.75 & 7725 & 25118.86 & 13.53 & 5.16 & 0.62 \\ 
  CoRoT-102788412 & 2.15 & 41.91 & 39.77 & 4.39 &  10 & 0.40 & 6.25 & 4.01 & 8000 & 8413.95 & 13.38 & 0.16 & 0.63 \\ 
  CoRoT-102688713 & 2.61 & 40.31 & 37.69 & 25.33 &  40 & 1.25 & 2.50 & 14.96 & 7300 & 14125.38 & 14.41 & 1.57 & 0.73 \\ 
  CoRoT-102589546 & 5.25 & 41.03 & 35.78 & 16.27 &  35 & 1.41 & 2.55 & 25.31 & 7250 & 5011.87 & 12.02 & 1.98 & 0.80 \\ 
  CoRoT-102690176 & 2.48 & 30.42 & 27.94 & 12.49 &  35 & 6.04 & 4.39 & 12.38 & 7425 & 3349.65 & 14.11 & 36.49 & 0.78 \\ 
  CoRoT-102790482 & 2.72 & 32.70 & 29.98 & 13.16 &  48 & 1.03 & 2.36 & 12.18 & 7225 & 2985.38 & 13.41 & 1.07 & 0.84 \\ 
  CoRoT-102591062 & 1.33 & 37.73 & 36.39 & 29.20 &  10 & 1.14 & 6.94 & 28.94 & 7600 & 4466.84 & 14.25 & 1.29 & 0.73 \\ 
  CoRoT-102691322 & 2.22 & 75.85 & 73.63 & 58.12 &  18 & 1.04 & 3.50 & 51.16 & 7650 & 11220.18 & 13.25 & 1.08 & 0.67 \\ 
  CoRoT-102691789 & 1.80 & 38.65 & 36.85 & 25.87 &  20 & 0.59 & 6.25 & 30.11 & 7800 & 5623.41 & 13.92 & 0.34 & 0.68 \\ 
  CoRoT-102794872 & 2.17 & 68.85 & 66.68 & 38.37 &  58 & 0.98 & 1.71 & 33.77 & 7575 & 14125.38 & 11.77 & 0.97 & 0.67 \\ 
  CoRoT-102596121 & 2.40 & 53.55 & 51.15 & 44.45 &  33 & 1.58 & 2.56 & 37.82 & 7700 & 10000.00 & 13.41 & 2.49 & 0.67 \\ 
  CoRoT-102598868 & 1.68 & 41.92 & 40.24 & 26.03 &  26 & 1.41 & 2.56 & 30.58 & 7750 & 7943.28 & 13.15 & 1.99 & 0.67 \\
  KIC-9776474 & 0.00 & 97.90 & 97.90 & 0.01 & 110 & 1.85 & -- & 19.47 & -- & -- & -- & 3.41 & -- \\ 
  KIC-11402951 & 0.01 & 51.31 & 51.30 & 23.85 & 202 & 0.88 & -- & 25.79 & 7290.00 & 3162.28 & 8.12 & 0.78 & 0.82 \\ 
  KIC-1718594 & 0.04 & 76.14 & 76.10 & 41.57 &  61 & 2.53 & -- & 6.12 & 7491.10 & 10118.12 & 10.31 & 6.40 & 0.71 \\ 
  KIC-2303365 & 0.07 & 22.39 & 22.32 & 14.81 &  76 & 7.39 & -- & 0.68 & 7280.80 & 5085.11 & 11.34 & 54.58 & 0.79 \\ 
  KIC-2439660 & 0.05 & 82.60 & 82.55 & 42.97 &  36 & 1.51 & -- & 42.40 & 7987.80 & 9906.04 & 11.55 & 2.29 & 0.62 \\ 
  KIC-2571868 & 0.04 & 60.16 & 60.12 & 20.55 & 209 & 1.92 & -- & 21.21 & 8120.00 & -- & 8.67 & 3.69 & -- \\ 
  KIC-2987660 & 0.07 & 54.74 & 54.67 & 15.05 & 303 & 5.23 & 3.97 & 14.71 & 7314.84 & 4037.89 & 7.93 & 27.35 & 0.79 \\ 
  KIC-3217554 & 0.03 & 47.75 & 47.72 & 5.60 & 173 & 2.43 & 2.46 & 7.63 & 7825.60 & 4904.56 & 9.56 & 5.89 & 0.68 \\ 
  KIC-3219256 & 0.01 & 89.65 & 89.64 & 17.86 & 606 & 1.45 & 2.89 & 20.79 & 7508.80 & 4090.27 & 8.31 & 2.10 & 0.75 \\ 
  KIC-3347643 & 0.12 & 95.67 & 95.55 & 29.69 & 284 & 1.04 & 5.01 & 22.62 & 8560.00 & 12761.45 & 7.69 & 1.09 & 0.53 \\ 
  KIC-3425802 & 0.07 & 60.88 & 60.81 & 0.07 &  36 & 1.26 & -- & 0.69 & 8044.90 & 12511.23 & 11.11 & 1.59 & 0.60 \\ 
  KIC-3429637 & 0.01 & 22.24 & 22.23 & 10.34 &  69 & 1.61 & -- & 0.28 & 7201.10 & 10092.53 & 7.72 & 2.58 & 0.76 \\ 
  KIC-3440495 & 0.03 & 26.14 & 26.11 & 0.03 &  29 & 0.38 & -- & 4.95 & 7409.40 & 7981.78 & 10.51 & 0.14 & 0.73 \\ 
  KIC-3634384 & 0.10 & 62.49 & 62.40 & 11.65 & 181 & 1.62 & -- & 27.25 & 7482.40 & 7908.61 & 11.07 & 2.62 & 0.72 \\ 
  KIC-3760002 & 0.05 & 38.36 & 38.31 & 16.26 & 100 & 2.29 & 4.10 & 1.96 & 7027.60 & 9990.79 & 10.65 & 5.27 & 0.80 \\ 
  KIC-3761641 & 0.59 & 82.74 & 82.15 & 37.31 &  40 & 0.83 & -- & 36.08 & 8149.40 & 12595.05 & 10.84 & 0.68 & 0.59 \\ 
  KIC-3850810 & 0.34 & 47.70 & 47.36 & 14.46 &  96 & 1.95 & 2.68 & 7.98 & 7200.00 & 4074.84 & 10.10 & 3.79 & 0.82 \\ 
  KIC-3941283 & 0.03 & 77.93 & 77.90 & 46.10 & 130 & 1.32 & 6.39 & 43.33 & 7797.70 & 7981.78 & 10.24 & 1.75 & 0.66 \\ 
  KIC-4035667 & 0.14 & 76.59 & 76.45 & 46.38 &  37 & 1.12 & 6.13 & 47.31 & 8050.00 & 7921.36 & 10.05 & 1.26 & 0.62 \\ 
  KIC-4048494 & 0.09 & 45.53 & 45.44 & 14.36 & 108 & 4.72 & -- & 13.82 & 7626.00 & 10195.30 & 9.55 & 22.32 & 0.68 \\ 
  KIC-4077032 & 0.03 & 52.86 & 52.83 & 14.48 & 325 & 3.81 & -- & 14.03 & 7220.00 & 12761.45 & 9.71 & 14.54 & 0.75 \\ 
  KIC-4168574 & 0.04 & 33.93 & 33.88 & 7.99 &  65 & 1.51 & -- & 4.73 & -- & -- & 10.29 & 2.28 & -- \\ 
  KIC-4269337 & 0.04 & 64.39 & 64.35 & 35.22 &  64 & 2.34 & -- & 4.28 & -- & -- & 10.83 & 5.46 & -- \\ 
  KIC-4383117 & 0.04 & 45.08 & 45.04 & 26.26 &  16 & 0.40 & -- & 27.88 & -- & -- & 10.80 & 0.16 & -- \\ 
  KIC-4647763 & 0.03 & 43.47 & 43.44 & 23.72 & 121 & 4.07 & 3.67 & 5.78 & 6843.40 & 3695.73 & 10.89 & 16.52 & 0.91 \\ 
  KIC-4649476 & 0.03 & 53.73 & 53.70 & 20.70 & 190 & 1.56 & 5.01 & 9.94 & 8322.20 & 8253.92 & 9.33 & 2.44 & 0.58 \\ 
  KIC-4856630 & 0.04 & 43.37 & 43.33 & 19.58 &  93 & 2.67 & 4.58 & 1.94 & 7359.90 & 8044.52 & 11.15 & 7.15 & 0.74 \\ 
  KIC-4863077 & 0.03 & 62.73 & 62.70 & 20.99 & 232 & 2.42 & 2.55 & 21.11 & 7524.70 & 7959.76 & 11.07 & 5.84 & 0.71 \\ 
  KIC-4936524 & 0.03 & 54.89 & 54.86 & 28.04 &  24 & 2.47 & -- & 26.62 & 7470.60 & 6335.78 & 13.12 & 6.10 & 0.74 \\ 
  KIC-5080290 & 0.03 & 22.56 & 22.53 & 0.03 &  95 & 4.81 & 2.68 & 1.78 & -- & -- & 9.56 & 23.15 & -- \\ 
  KIC-5209712 & 0.00 & 74.59 & 74.58 & 14.05 & 109 & 0.95 & 2.89 & 34.81 & 8358.90 & 9956.35 & 11.20 & 0.89 & 0.57 \\ 
  KIC-5272673 & 0.04 & 46.03 & 45.99 & 16.81 & 133 & 10.26 & -- & 2.12 & -- & -- & 10.34 & 105.34 & -- \\ 
  KIC-5391416 & 0.02 & 56.53 & 56.51 & 9.62 & 148 & 2.78 & 2.64 & 3.00 & 8062.60 & 7992.82 & 10.19 & 7.70 & 0.62 \\ 
  KIC-5428254 & 0.28 & 73.90 & 73.62 & 19.16 & 182 & 2.81 & 5.05 & 28.49 & 7380.90 & 7837.91 & 10.46 & 7.88 & 0.74 \\ 
  KIC-5632093 & 0.08 & 89.13 & 89.04 & 47.23 &  53 & 1.36 & 6.48 & 43.10 & 8088.80 & 12612.47 & 10.88 & 1.84 & 0.60 \\ 
  KIC-5709664 & 0.05 & 41.68 & 41.63 & 19.44 &  35 & 0.91 & -- & 0.79 & 7201.10 & 10092.53 & 11.53 & 0.83 & 0.76 \\ 
  KIC-5785707 & 0.03 & 74.66 & 74.63 & 0.03 & 221 & 2.83 & -- & 3.51 & -- & -- & 9.03 & 7.98 & -- \\ 
  KIC-6586052 & 0.03 & 68.41 & 68.38 & 20.88 &  97 & 4.80 & 3.80 & 23.67 & 7515.90 & 12691.12 & 11.14 & 23.01 & 0.69 \\ 
  KIC-6629106 & 0.74 & 27.43 & 26.69 & 16.94 &  34 & 2.78 & -- & 15.17 & 7078.30 & 9896.92 & 10.08 & 7.72 & 0.79 \\ 
  KIC-6668729 & 0.04 & 56.22 & 56.18 & 21.16 & 321 & 2.05 & 2.76 & 17.45 & -- & -- & 8.57 & 4.21 & -- \\ 
  KIC-6790335 & 0.04 & 71.13 & 71.09 & 20.27 & 343 & 3.41 & 2.76 & 13.30 & -- & -- & 10.51 & 11.63 & -- \\ 
  KIC-6804821 & 0.23 & 31.40 & 31.17 & 14.52 &  29 & 0.82 & 2.59 & 1.03 & 7479.30 & 4945.38 & 10.51 & 0.67 & 0.75 \\ 
  KIC-6865077 & 0.02 & 53.24 & 53.22 & 24.17 &  76 & 2.49 & 7.82 & 9.64 & -- & -- & 9.76 & 6.21 & -- \\ 
  KIC-6937758 & 0.52 & 52.47 & 51.95 & 20.08 & 126 & 5.65 & 4.06 & 20.37 & 7760.00 & -- & 9.72 & 31.92 & -- \\ 
  KIC-6965789 & 0.02 & 61.70 & 61.68 & 16.32 & 290 & 3.22 & 3.07 & 2.14 & -- & -- & 10.06 & 10.39 & -- \\ 
  KIC-7106205 & 0.04 & 22.45 & 22.42 & 13.39 &  44 & 4.91 & 2.89 & 0.51 & 6905.80 & 5407.54 & 11.38 & 24.13 & 0.87 \\ 
  KIC-7217483 & 0.34 & 28.36 & 28.02 & 13.93 &  70 & 4.80 & 4.67 & 2.18 & 6944.30 & 7967.10 & 10.54 & 23.07 & 0.84 \\ 
  KIC-7265427 & 0.05 & 53.45 & 53.40 & 24.14 & 154 & 1.99 & 2.72 & 29.55 & 7564.40 & 12522.76 & 11.32 & 3.96 & 0.68 \\ 
  KIC-7450284 & 0.21 & 51.30 & 51.09 & 19.73 & 176 & 3.20 & 3.37 & 21.29 & 7928.90 & 5053.59 & 10.38 & 10.24 & 0.66 \\ 
  KIC-7548479 & 0.03 & 73.26 & 73.24 & 21.71 & 136 & 1.73 & 3.54 & 1.25 & 7494.10 & 7809.08 & 8.40 & 3.00 & 0.72 \\ 
  KIC-7583939 & 0.07 & 93.59 & 93.53 & 23.16 & 395 & 1.37 & -- & 31.84 & 8170.00 & 8148.90 & 9.64 & 1.87 & 0.60 \\ 
  KIC-7697795 & 0.10 & 42.02 & 41.92 & 17.49 & 150 & 2.74 & -- & 18.09 & -- & -- & 10.78 & 7.53 & -- \\ 
  KIC-7773133 & 0.05 & 38.28 & 38.23 & 5.83 & 412 & 23.53 & 4.10 & 7.00 & 6630.63 & 3042.22 & 10.93 & 553.66 & 0.99 \\ 
  KIC-7834612 & 0.10 & 42.46 & 42.36 & 8.53 & 169 & 4.82 & 4.15 & 12.30 & 7455.20 & 5009.56 & 10.25 & 23.24 & 0.75 \\ 
  KIC-7842286 & 0.10 & 70.32 & 70.21 & 26.42 & 123 & 2.89 & 3.93 & 31.84 & 7820.00 & 6212.98 & 9.93 & 8.37 & 0.67 \\ 
  KIC-7842621 & 0.07 & 45.42 & 45.35 & 0.07 &  41 & 1.23 & -- & 29.55 & -- & -- & 10.84 & 1.52 & -- \\ 
  KIC-8103917 & 0.05 & 38.65 & 38.60 & 17.75 &  40 & 2.00 & -- & 19.00 & 7125.80 & 16043.53 & 11.81 & 3.99 & 0.76 \\ 
  KIC-8245366 & 0.06 & 43.49 & 43.44 & 11.94 & 296 & 28.97 & -- & 8.37 & -- & -- & 11.19 & 839.53 & -- \\ 
  KIC-8264546 & 0.03 & 38.37 & 38.33 & 24.99 &  37 & 1.19 & 3.37 & 25.30 & 7628.70 & 10006.91 & 13.41 & 1.41 & 0.68 \\ 
  KIC-8352420 & 0.01 & 13.13 & 13.13 & 9.27 &  31 & 5.40 & -- & 0.48 & -- & -- & 12.03 & 29.14 & -- \\ 
  KIC-8415752 & 0.11 & 72.10 & 71.99 & 37.76 &  75 & 0.42 & 5.18 & 20.05 & 7774.30 & 9928.87 & 10.66 & 0.17 & 0.66 \\ 
  KIC-8429756 & 0.03 & 55.56 & 55.54 & 27.75 &  66 & 0.74 & -- & 26.61 & 7362.10 & 5045.45 & 10.61 & 0.54 & 0.77 \\ 
  KIC-8446738 & 0.03 & 67.89 & 67.86 & 38.25 &  48 & 0.99 & 5.18 & 36.59 & 7153.80 & 7930.49 & 10.92 & 0.99 & 0.79 \\ 
  KIC-8459354 & 0.00 & 58.13 & 58.12 & 19.47 & 240 & 3.91 & 2.72 & 19.97 & 7429.76 & 3989.99 & 11.12 & 15.30 & 0.77 \\ 
  KIC-8525286 & 0.06 & 52.88 & 52.82 & 0.06 &  85 & 5.14 & 4.45 & 1.54 & 7344.80 & 15995.58 & -- & 26.41 & 0.71 \\ 
  KIC-8560996 & 0.09 & 54.34 & 54.26 & 20.94 &  29 & 0.95 & -- & 0.71 & -- & -- & 10.46 & 0.90 & -- \\ 
  KIC-8565229 & 0.04 & 70.74 & 70.69 & 22.54 & 163 & 3.92 & 3.02 & 22.18 & 7745.30 & 8044.52 & 10.98 & 15.33 & 0.67 \\ 
  KIC-8623953 & 0.15 & 76.78 & 76.63 & 27.26 & 238 & 14.92 & 4.92 & 37.49 & -- & -- & 9.32 & 222.46 & -- \\ 
  KIC-8695156 & 0.09 & 17.71 & 17.62 & 5.77 &  44 & 3.36 & -- & 3.92 & -- & -- & 10.74 & 11.29 & -- \\ 
  KIC-8717065 & 0.46 & 66.33 & 65.87 & 34.98 &  46 & 0.87 & 4.71 & 35.74 & 7435.50 & 6316.84 & 10.94 & 0.75 & 0.74 \\ 
  KIC-8750029 & 0.29 & 30.61 & 30.32 & 22.44 &  14 & 0.59 & -- & 4.97 & 7540.00 & 4978.52 & 9.62 & 0.35 & 0.74 \\ 
  KIC-8827821 & 0.06 & 47.61 & 47.55 & 17.71 &  97 & 1.70 & 4.28 & 23.79 & 7389.20 & 5046.61 & 11.14 & 2.90 & 0.77 \\ 
  KIC-8869892 & 0.03 & 17.72 & 17.69 & 7.70 &  43 & 5.50 & 2.55 & 0.81 & -- & -- & -- & 30.21 & -- \\ 
  KIC-8881697 & 0.99 & 58.52 & 57.53 & 16.56 &  73 & 1.88 & 5.23 & 32.03 & 7728.50 & 7961.59 & 10.50 & 3.54 & 0.68 \\ 
  KIC-8933391 & 0.03 & 17.68 & 17.65 & 6.71 &  35 & 0.19 & 3.28 & 1.57 & 7520.00 & 4997.09 & 8.85 & 0.04 & 0.74 \\ 
  KIC-9020199 & 1.26 & 22.91 & 21.66 & 7.08 &  56 & 4.75 & 2.55 & 7.79 & 6548.70 & 10027.67 & 8.86 & 22.55 & 0.93 \\ 
  KIC-9108615 & 0.07 & 13.52 & 13.45 & 0.07 &  21 & 1.43 & 2.94 & 0.30 & 6701.00 & 6401.77 & 11.35 & 2.03 & 0.91 \\ 
  KIC-9138872 & 0.06 & 46.89 & 46.82 & 20.71 & 165 & 2.86 & -- & 18.99 & 7493.50 & 9931.16 & 10.03 & 8.15 & 0.71 \\ 
  KIC-9143785 & 0.36 & 49.46 & 49.10 & 11.80 &  72 & 2.39 & 3.67 & 23.84 & 7603.10 & 19984.81 & 11.93 & 5.69 & 0.65 \\ 
  KIC-9156808 & 0.38 & 57.86 & 57.48 & 21.28 &  62 & 2.52 & -- & 1.26 & 7065.90 & 7950.60 & 11.29 & 6.35 & 0.81 \\ 
  KIC-9201644 & 0.01 & 49.81 & 49.80 & 14.72 & 162 & 2.09 & 4.36 & 12.83 & -- & -- & 10.99 & 4.37 & -- \\ 
  KIC-9210037 & 0.45 & 30.27 & 29.82 & 9.28 &  70 & 7.72 & 4.49 & 9.61 & -- & -- & -- & 59.63 & -- \\ 
  KIC-9229318 & 0.02 & 56.44 & 56.42 & 6.24 & 300 & 5.24 & -- & 18.82 & 7136.50 & 4955.64 & 9.59 & 27.46 & 0.82 \\ 
  KIC-9246481 & 1.90 & 81.20 & 79.30 & 40.87 &  64 & 0.46 & -- & 18.03 & 8180.00 & -- & 9.04 & 0.21 & -- \\ 
  KIC-9267042 & 0.65 & 93.99 & 93.34 & 24.66 &  85 & 9.10 & -- & 26.74 & -- & -- & -- & 82.76 & -- \\ 
  KIC-9291618 & 0.03 & 64.99 & 64.96 & 10.32 & 286 & 3.35 & 4.92 & 4.67 & 7970.00 & -- & 9.66 & 11.24 & -- \\ 
  KIC-9353572 & 0.01 & 24.28 & 24.27 & 13.39 &  33 & 1.89 & -- & 0.61 & 7186.20 & 9988.49 & 10.51 & 3.57 & 0.77 \\ 
  KIC-9395246 & 0.20 & 22.93 & 22.73 & 7.70 &  54 & 4.98 & -- & 7.41 & -- & -- & 11.77 & 24.79 & -- \\ 
  KIC-9450940 & 0.00 & 87.85 & 87.84 & 30.00 &  61 & 5.35 & -- & 42.61 & 8206.00 & 15995.58 & -- & 28.62 & 0.57 \\ 
  KIC-9453075 & 0.06 & 39.35 & 39.28 & 19.31 &  20 & 2.42 & -- & 0.68 & 8060.00 & 9972.41 & 11.74 & 5.85 & 0.61 \\ 
  KIC-9533449 & 0.03 & 24.25 & 24.22 & 12.52 &  70 & 3.20 & -- & 7.06 & -- & -- & 11.33 & 10.23 & -- \\ 
  KIC-9551281 & 0.04 & 45.53 & 45.49 & 21.48 & 150 & 1.58 & 3.54 & 24.56 & 7455.20 & 5009.56 & 10.45 & 2.48 & 0.75 \\ 
  KIC-9580794 & 0.81 & 43.23 & 42.42 & 10.19 &  76 & 2.93 & 5.53 & 36.06 & 7297.50 & 12491.08 & 11.27 & 8.58 & 0.73 \\ 
  KIC-9642894 & 0.13 & 29.67 & 29.54 & 14.68 &  88 & 8.38 & -- & 5.59 & 7053.60 & 12714.52 & 11.27 & 70.29 & 0.78 \\ 
  KIC-9655055 & 0.05 & 25.10 & 25.05 & 7.84 &  71 & 2.28 & -- & 7.65 & -- & -- & 11.35 & 5.20 & -- \\ 
  KIC-9655114 & 0.01 & 97.91 & 97.90 & 20.57 & 303 & 3.67 & -- & 25.38 & 7409.40 & 7981.78 & 12.03 & 13.48 & 0.73 \\ 
  KIC-9655177 & 0.22 & 28.54 & 28.32 & 8.51 & 126 & 7.35 & -- & 7.23 & 6896.40 & 8029.71 & 11.07 & 54.01 & 0.85 \\ 
  KIC-9655422 & 0.09 & 31.92 & 31.83 & 5.13 & 125 & 5.69 & 4.23 & 6.79 & -- & -- & 11.25 & 32.38 & -- \\ 
  KIC-9655514 & 0.00 & 97.96 & 97.95 & 15.18 & 252 & 3.57 & -- & 15.01 & -- & -- & 11.88 & 12.71 & -- \\ 
  KIC-9673293 & 0.05 & 50.13 & 50.08 & 29.12 &  25 & 0.26 & -- & 40.71 & 8227.50 & 7957.93 & 10.56 & 0.07 & 0.60 \\ 
  KIC-9693282 & 0.05 & 54.41 & 54.36 & 8.68 & 339 & 4.18 & -- & 15.57 & -- & -- & 10.40 & 17.50 & -- \\ 
  KIC-9699950 & 0.09 & 38.29 & 38.20 & 17.01 &  31 & 5.60 & 3.93 & 17.65 & -- & -- & -- & 31.40 & -- \\ 
  KIC-9700145 & 0.06 & 36.17 & 36.10 & 13.06 &  53 & 4.45 & 5.05 & 0.78 & 7588.00 & 9935.74 & -- & 19.82 & 0.69 \\ 
  KIC-9700322 & 0.00 & 24.29 & 24.28 & 12.57 &  40 & 29.44 & 2.89 & 0.40 & 6701.40 & 5061.74 & 12.61 & 866.59 & 0.93 \\ 
  KIC-9762713 & 0.18 & 28.41 & 28.24 & 13.86 &  66 & 4.02 & -- & 8.81 & 6965.30 & 12496.83 & 11.00 & 16.14 & 0.81 \\ 
  KIC-9773512 & 0.10 & 21.11 & 21.01 & 9.21 &  30 & 2.30 & 4.02 & 3.09 & -- & -- & 10.02 & 5.31 & -- \\ 
  KIC-9812351 & 0.70 & 93.56 & 92.86 & 18.58 & 435 & 4.59 & -- & 17.69 & -- & -- & 7.90 & 21.07 & -- \\ 
  KIC-9818269 & 0.15 & 36.94 & 36.79 & 19.18 &  32 & 1.92 & -- & 21.02 & -- & -- & -- & 3.68 & -- \\ 
  KIC-9836020 & 0.01 & 22.34 & 22.33 & 0.01 &  29 & 1.49 & -- & 19.97 & 7480.70 & 12670.68 & 11.95 & 2.23 & 0.70 \\ 
  KIC-9845907 & 1.02 & 88.99 & 87.97 & 17.60 &  88 & 35.78 & 3.46 & 30.20 & 7945.40 & 12485.33 & 11.42 & 1280.33 & 0.62 \\ 
  KIC-10002897 & 0.25 & 36.42 & 36.17 & 15.39 &  40 & 4.12 & -- & 23.65 & 7489.85 & 5110.58 & 12.87 & 17.01 & 0.74 \\ 
  KIC-10056297 & 0.25 & 29.14 & 28.89 & 9.56 & 114 & 33.40 & -- & 10.03 & 7156.40 & 9995.40 & -- & 1115.41 & 0.77 \\ 
  KIC-10134800 & 1.34 & 53.00 & 51.66 & 27.06 &  74 & 3.54 & -- & 13.80 & -- & -- & 11.60 & 12.52 & -- \\ 
  KIC-10253943 & 0.07 & 90.48 & 90.41 & 25.21 & 326 & 7.03 & 4.06 & 40.64 & -- & -- & 10.18 & 49.38 & -- \\ 
  KIC-10273384 & 0.01 & 83.47 & 83.46 & 8.22 & 241 & 18.86 & 7.78 & 7.94 & -- & -- & 11.68 & 355.68 & -- \\ 
  KIC-10289211 & 0.16 & 68.88 & 68.71 & 19.81 & 273 & 2.09 & 3.28 & 24.52 & -- & -- & 10.26 & 4.37 & -- \\ 
  KIC-10355055 & 0.20 & 61.55 & 61.35 & 22.08 & 132 & 3.43 & -- & 28.11 & 8420.00 & -- & 9.34 & 11.79 & -- \\ 
  KIC-10448764 & 0.09 & 37.62 & 37.53 & 9.59 & 212 & 10.78 & 4.23 & 10.29 & 7405.90 & 9977.00 & 10.90 & 116.13 & 0.72 \\ 
  KIC-10451090 & 0.09 & 76.75 & 76.67 & 38.38 & 121 & 2.34 & 2.68 & 34.72 & 7642.40 & 5683.29 & 9.12 & 5.48 & 0.71 \\ 
  KIC-10484808 & 0.05 & 62.62 & 62.57 & 27.12 &  58 & 1.15 & 5.40 & 33.09 & 8142.10 & 10118.12 & 10.52 & 1.32 & 0.60 \\ 
  KIC-10533616 & 0.09 & 58.50 & 58.40 & 0.09 &  18 & 0.16 & 4.45 & 1.73 & 8260.00 & -- & 9.57 & 0.02 & -- \\ 
  KIC-10549371 & 0.20 & 28.63 & 28.44 & 13.88 &  99 & 4.15 & -- & 10.85 & 7300.00 & 10073.95 & 9.43 & 17.22 & 0.74 \\ 
  KIC-10590857 & 0.78 & 47.94 & 47.16 & 20.93 &  91 & 2.69 & -- & 24.88 & 7470.00 & 6258.93 & 9.99 & 7.23 & 0.74 \\ 
  KIC-10604429 & 0.02 & 43.95 & 43.94 & 0.02 & 262 & 0.98 & -- & 21.05 & -- & -- & 9.93 & 0.95 & -- \\ 
  KIC-10615125 & 0.04 & 37.42 & 37.38 & 9.66 & 174 & 5.08 & 2.98 & 10.09 & -- & -- & 10.34 & 25.84 & -- \\ 
  KIC-10658802 & 0.36 & 39.31 & 38.95 & 8.16 &  63 & 3.02 & 6.61 & 8.02 & 7480.70 & 12670.68 & 12.04 & 9.10 & 0.70 \\ 
  KIC-10684673 & 0.03 & 11.50 & 11.47 & 10.39 &  44 & 7.44 & -- & 0.81 & 7116.00 & 8046.37 & 11.19 & 55.42 & 0.80 \\ 
  KIC-10686752 & 0.18 & 72.24 & 72.07 & 40.47 &  52 & 1.02 & -- & 37.47 & -- & -- & 11.32 & 1.04 & -- \\ 
  KIC-10713398 & 1.74 & 45.29 & 43.55 & 15.35 &  83 & 3.13 & -- & 16.44 & -- & -- & 11.21 & 9.82 & -- \\ 
  KIC-10717871 & 0.04 & 48.03 & 47.99 & 12.22 & 199 & 3.79 & 2.55 & 15.48 & 7290.00 & 3162.28 & 10.52 & 14.37 & 0.82 \\ 
  KIC-10775968 & 0.05 & 40.37 & 40.32 & 21.55 &  32 & 0.43 & -- & 3.50 & 7490.00 & 6309.57 & 10.60 & 0.19 & 0.73 \\ 
  KIC-10777903 & 0.03 & 48.35 & 48.32 & 15.86 & 280 & 4.51 & -- & 14.79 & 7320.00 & 3162.28 & 10.12 & 20.33 & 0.81 \\ 
  KIC-10788451 & 0.47 & 50.56 & 50.10 & 23.16 & 120 & 4.26 & 3.93 & 18.57 & 8790.00 & 10000.00 & 11.15 & 18.19 & 0.51 \\ 
  KIC-10813970 & 0.04 & 45.58 & 45.54 & 9.61 & 169 & 5.25 & -- & 8.24 & 7420.00 & 3981.07 & 11.06 & 27.56 & 0.77 \\ 
  KIC-10815466 & 0.09 & 88.61 & 88.53 & 36.71 & 256 & 3.42 & 2.76 & 54.11 & 8310.00 & 7943.28 & 11.10 & 11.69 & 0.58 \\ 
  KIC-10977859 & 0.04 & 82.53 & 82.48 & 59.97 & 102 & 2.89 & -- & 1.84 & 8400.00 & 4073.80 & 8.76 & 8.33 & 0.60 \\ 
  KIC-10988009 & 0.06 & 72.63 & 72.57 & 12.96 & 218 & 5.51 & -- & 16.21 & 7320.00 & 10000.00 & 10.10 & 30.40 & 0.74 \\ 
  KIC-11013201 & 0.07 & 58.37 & 58.31 & 33.42 &  36 & 0.72 & -- & 1.76 & 7740.00 & 3162.28 & 9.31 & 0.52 & 0.72 \\ 
  KIC-11090405 & 0.04 & 38.11 & 38.07 & 14.37 &  69 & -- & -- & 23.81 & 7900.00 & 5128.61 & 9.53 & 0.99 & 0.67 \\ 
  KIC-11125764 & 0.03 & 66.00 & 65.97 & 38.33 &  77 & 1.87 & -- & 33.04 & 7980.00 & 7943.28 & 10.94 & 3.49 & 0.63 \\ 
  KIC-11127190 & 0.15 & 57.30 & 57.15 & 18.73 &  52 & 4.36 & -- & 18.76 & 7630.00 & 5011.87 & 11.42 & 19.04 & 0.72 \\ 
  KIC-11183539 & 0.19 & 59.21 & 59.03 & 22.66 & 198 & 2.29 & 3.28 & 27.18 & 7470.00 & 3162.28 & 10.76 & 5.26 & 0.78 \\ 
  KIC-11340713 & 0.21 & 20.05 & 19.84 & 10.15 &  53 & 11.33 & 2.94 & 3.40 & -- & -- & -- & 128.43 & -- \\ 
  KIC-11395392 & 0.04 & 46.17 & 46.14 & 25.19 & 110 & 6.58 & 3.02 & 22.27 & -- & -- & -- & 43.26 & -- \\ 
  KIC-11497012 & 0.24 & 54.03 & 53.79 & 24.11 & 198 & 1.67 & 2.12 & 23.24 & 7530.00 & 7943.28 & 9.66 & 2.79 & 0.71 \\ 
  KIC-11661993 & 0.19 & 44.93 & 44.74 & 14.03 & 247 & 3.48 & 2.64 & 16.23 & 7200.00 & 7943.28 & 9.33 & 12.10 & 0.78 \\ 
  KIC-11700370 & 0.05 & 74.35 & 74.31 & 0.05 &  49 & 0.16 & -- & 49.01 & 8290.00 & 10000.00 & 10.67 & 0.02 & 0.58 \\ 
  KIC-11754974 & 0.08 & 86.29 & 86.21 & 16.35 & 107 & 57.66 & 5.40 & 20.95 & -- & -- & 12.49 & 3324.53 & -- \\ 
  KIC-11821140 & 2.07 & 61.30 & 59.23 & 38.79 &  35 & 0.33 & -- & 38.64 & 7700.00 & 7943.28 & 10.01 & 0.11 & 0.68 \\ 
  KIC-11874676 & 0.03 & 80.97 & 80.94 & 13.58 &  67 & 1.82 & -- & 15.16 & 8220.00 & 10000.00 & 10.10 & 3.32 & 0.59 \\ 
  KIC-12020590 & 0.20 & 36.24 & 36.04 & 22.21 &  20 & 0.45 & -- & 31.82 & 7950.00 & 5011.87 & 9.91 & 0.21 & 0.66 \\ 
  KIC-12068180 & 0.07 & 52.81 & 52.74 & 20.30 & 229 & 2.25 & -- & 20.84 & 7460.00 & 6309.57 & 10.36 & 5.07 & 0.74 \\ 
  KIC-12353648 & 0.06 & 27.28 & 27.22 & 7.82 &  84 & 3.93 & -- & 1.75 & 7190.00 & 3981.07 & 9.48 & 15.43 & 0.82 \\ 
    \hline
 \end{longtable}
\end{landscape}







\bsp	
\label{lastpage}
\end{document}